\documentclass[nofootinbib,twocolumn,preprintnumbers]{revtex4-1}
\pdfoutput=1

\usepackage{amsmath,amssymb,graphicx,booktabs,bm,psfrag,color}

\begin{document}

\title{Flavor Anomalies, the Diphoton Excess and a Dark Matter Candidate}

\preprint{MITP/15-120}
%arXiv:1512.06828
%v1: December 21, 2015
%v2: January 21, 2016
%v3: May 24, 2016

\author{Martin Bauer$^a$}
\author{Matthias Neubert$^{b,c}$}

\affiliation{$^a$Institut f\"ur Theoretische Physik, Universit\"at Heidelberg, Philosophenweg 16, 69120 Heidelberg, Germany\\
${}^b$PRISMA Cluster of Excellence {\em\&} MITP, Johannes Gutenberg University, 55099 Mainz, Germany\\
${}^c$Department of Physics {\em\&} LEPP, Cornell University, Ithaca, NY 14853, U.S.A.}

\begin{abstract}  
We argue that the diphoton excess recently reported by ATLAS and CMS can be explained, along with several anomalies seen in the flavor sector, in models where a Standard-Model singlet scalar resonance with mass $M_S\approx 750$\,GeV is produced in gluon fusion via loops containing a scalar color-triplet leptoquark $\phi$. For a leptoquark mass $M_\phi\lesssim 1$\,TeV, the production cross section is naturally in the 10\,fb range. A large $S\to\gamma\gamma$ branching ratio can be obtained by coupling the scalar $S$ to new color-singlet fermions $\chi$ with electroweak scale masses, which can be part of an $SU(2)_L$ multiplet, whose neutral component has the right mass and quantum numbers to be a dark matter candidate. Our model reveals a connection between flavor anomalies, the nature of dark matter and a new scalar, which acts as a mediator to the dark sector. The loop-mediated decay $S\to\tau^+\tau^-$ could be a striking signature of this model.
\end{abstract}
\maketitle

\section{Introduction}

Recently, the ATLAS and CMS collaborations have reported an excess in the $\sqrt{s}=13$\,TeV diphoton spectrum at a mass of $M_{\gamma\gamma}\approx 750$\,GeV with local significance of $3.6\,\sigma$ (ATLAS) and $2.6\,\sigma$ (CMS) \cite{750GeVExp}. Taking into account the look-elsewhere effect, the global significance of the excess is reduced to $1.9\,\sigma$ (ATLAS) and $1.2\,\sigma$ (CMS). If confirmed by future data, this would be a tantalizing first direct signal of physics beyond the Standard Model (SM) with far-reaching impact on our understanding of Nature. The absence of any significant excess in the dijet, massive diboson and $t\bar t$ spectra provides several important hints on the properties of a possible resonance $S$. This new resonance is very likely neutral under the SM gauge group, carries spin 0 or 2 and its decays into massive electroweak gauge bosons and SM fermions need to be sufficiently suppressed. In particular, couplings to light fermions are strongly constrained by dijet searches and precision flavor observables. The observed cross section $\sigma(pp\to S\to\gamma\gamma)=(4.4\pm 1.1)$\,fb \cite{Buttazzo:2015txu} thus implies a sizable coupling to gluons and a large branching ratio into the diphoton final state. Exceptions are models in which the resonance is produced through light-by-light scattering, for which its coupling to gluons can be negligible, while an enormous effective coupling to photons is required \cite{Fichet:2015vvy,Csaki:2015vek}. For a neutral scalar, the couplings to gluons and photons need to be loop induced. However, production and decay through SM particles are problematic for a scalar mass $M_S\approx 750$\,GeV, because tree-level decays of the new resonance into these SM particles would completely dominate the branching fractions. Therefore, the SM needs to be extended beyond the new scalar singlet in order to explain the excess. This potentially opens the door to a whole new sector of physics.

Many of the recent studies of the diphoton resonance make the economic assumption of a single new particle mediating both the gluon fusion and the diphoton loop amplitudes \cite{Harigaya:2015ezk,Pilaftsis:2015ycr,Franceschini:2015kwy,DiChiara:2015vdm,Molinaro:2015cwg,Gupta:2015zzs,Low:2015qep,Ellis:2015oso,McDermott:2015sck,Bai:2015nbs,Aloni:2015mxa,Falkowski:2015swt,Agrawal:2015dbf,Curtin:2015jcv,Chao:2015ttq,No:2015bsn,Martinez:2015kmn,Kobakhidze:2015ldh,Matsuzaki:2015che,Angelescu:2015uiz,Mambrini:2015wyu,Backovic:2015fnp,Knapen:2015dap}. In most cases, the new particle is solely motivated by the diphoton excess. In this letter we pursue a different approach. A significant part of the LHC Run-I legacy consists of the observation of a number of intriguing anomalies in the flavor sector. This includes a 2.6$\,\sigma$ deviation of the ratio $R_K=\Gamma(B\to K\mu^+\mu^-)/\Gamma(B\to K e^+ e^-)$ from~1 \cite{Aaij:2014ora}, a discrepancy in some angular observables in $B\to K^*\mu^+\mu^-$ decays \cite{Aaij:2013qta}, a deviation of the $B_s\to\phi\,\mu^+\mu^-$ branching ratio from its SM value \cite{Aaij:2015esa} and a reinforcement of a previously observed anomaly in $\bar B\to D^*\tau\bar\nu$ decays \cite{Lees:2012xj,Lees:2013uzd,Matyja:2007kt,Adachi:2009qg,Bozek:2010xy,Aaij:2015yra}. Several authors have proposed that scalar color-triplet leptoquarks can explain one or more of these anomalies \cite{Fajfer:2012jt,Tanaka:2012nw,Dorsner:2013tla,Sakaki:2013bfa,Freytsis:2015qca,Hiller:2014yaa}; indeed, we have recently shown that a leptoquark with mass $M_\phi\lesssim 1$\,TeV, transforming as $(\bm{3},\bm{1},-\frac13)$ under the SM gauge group, can explain these anomalies in a natural way, while at the same time accounting for the anomalous magnetic moment of the muon \cite{Bauer:2015knc}. Here we argue that TeV-scale scalar leptoquarks can naturally account for the observed production rate of the new resonance $S$, with which they can interact through a portal coupling
\begin{equation}\label{eq:ourop}
   {\cal L} = \kappa_{\phi S} S\,\phi^\dagger\phi \,,
\end{equation}
with a dimensionful parameter $\kappa_{\phi S}$. The corresponding Higgs portal is assumed to be very small. Interestingly, a gluon-induced production cross section $\sigma(gg\to S)$ of order 10\,fb is generated in these models for leptoquark masses $M_\phi\lesssim 1$\,TeV and natural values of $\kappa_{\phi S}\equiv g_{\phi S} M_\phi$ with $g_{\phi S}=\mathcal{O}(1)$. However, these scenarios lead to a strongly suppressed diphoton branching ratio in the range $10^{-4}\!-\!10^{-3}$. We show that in an extension of these models by a color-neutral $SU(2)_L$ multiplet of fermions with mass $M_\chi\gtrsim M_S/2$, which contains charged states in addition to a neutral dark-matter candidate, the $S\to\gamma\gamma$ branching ratio can be greatly enhanced. Beyond the signatures characteristic for an extension including new color-neutral fermions, which set our model apart from the aforementioned ``one particle in the loop'' explanations, leptoquarks may further give rise to interesting loop-suppressed leptonic decay modes of the singlet $S$, which could provide smoking-gun signals of a flavor-motivated explanation of the diphoton excess.

\section{Production in gluon fusion}

\begin{figure}
\includegraphics[width=0.5\textwidth]{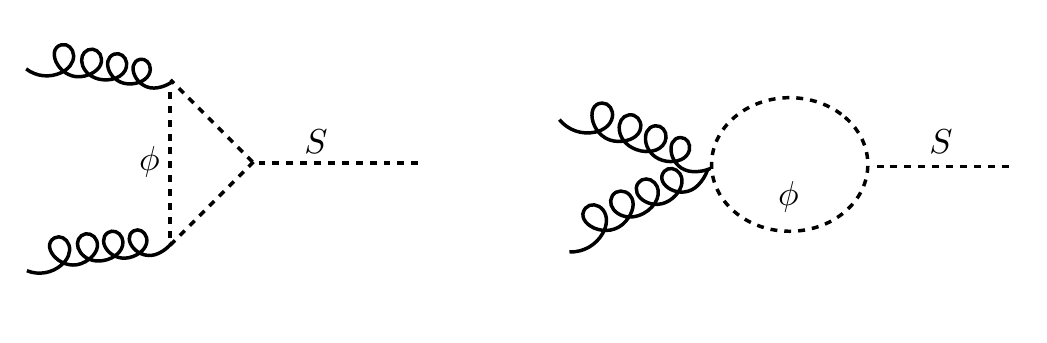}
\vspace{-8mm}
\caption{\label{fig:prod} 
Production of the scalar resonance $S$ in gluon fusion.}
\end{figure}

\begin{figure}[t!]
\vspace{2mm}
\includegraphics[width=0.4\textwidth]{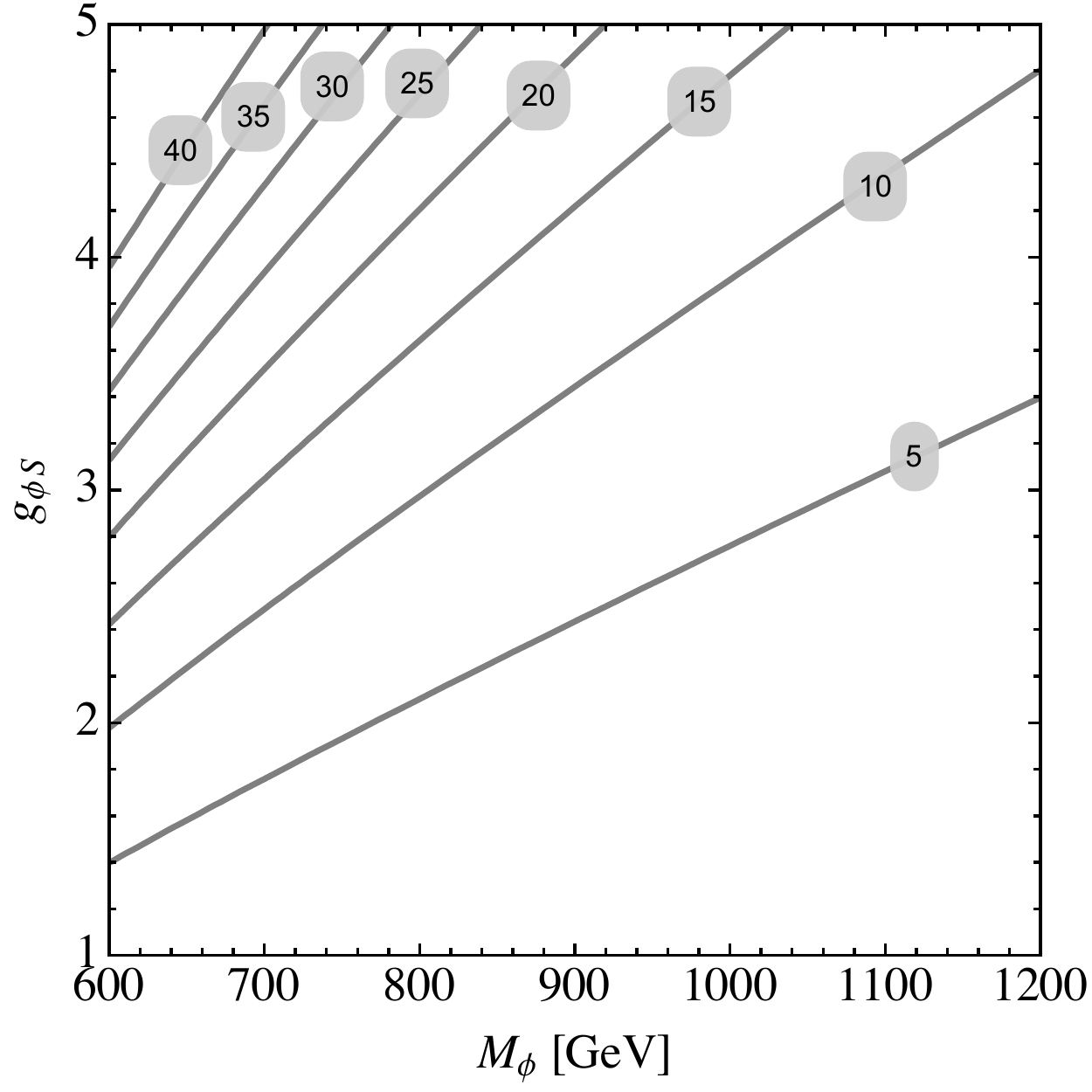}
\caption{\label{fig:production} 
Estimated $pp\to S$ production cross section (in fb) in gluon fusion at the LHC with $\sqrt{s}=13$\,TeV assuming $M_S=750$\,GeV, as a function of the leptoquark mass $M_\phi$ and the portal coupling $g_{\phi S}$.}
\end{figure}

Based on \eqref{eq:ourop} and the diagrams shown in Figure~\ref{fig:prod}, the gluon-fusion initiated production cross section of the scalar singlet $S$ reads
\begin{equation}
   \sigma(pp\to S) 
   = \frac{\pi}{s}\,\bigg[ \frac{\alpha_s}{192\pi}\,\frac{g_{\phi S} M_S}{M_\phi}\,A_0(\tau_\phi) \bigg]^2 
    K_{gg}\,f\hspace{-1.5mm}f_{gg}\big( M_S^2/s\big) \,,
\end{equation}
where $K_{gg}\approx 2$ accounts for higher-order QCD corrections when using $\alpha_s\equiv\alpha_s(M_S)\approx 0.092$, $\tau_\phi=4M_\phi^2/M_S^2$, 
\begin{equation}
   A_0(\tau) = 3\tau \big[ \tau\,f(\tau) - 1 \big] \,, \quad
   f(\tau) = \arcsin^2\Big( \frac{1}{\sqrt{\tau}} \Big) \,,
\end{equation}
with $A_0(\infty)=1$ the relevant loop function, and
\begin{equation}
   f\hspace{-1.5mm}f_{gg}(y) = \int_y^1\!\frac{dx}{x}\,f_{g/p}(x)\,f_{g/p}(y/x) 
\end{equation}
the gluon-gluon luminosity function. It is instructive to normalize the cross section to the cross section for the production of a hypothetical SM Higgs boson $h$ with mass $m_h=M_S$ and vacuum expectation value $\langle h\rangle=v$. We then obtain independently of the $K_{gg}$ factor
\begin{equation}\label{eq:sigmaratio}
   \frac{\sigma(pp\to S)}{\sigma(pp\to h)} 
   = \left( \frac{g_{\phi S}\,v}{8M_\phi} \right)^2
    \left| \frac{A_0(\tau_\phi)}{A_{1/2}(\tau_t)} \right|^2 ,
\end{equation}
where $\tau_t=4m_t^2/M_S^2$ and
\begin{equation}
   A_{1/2}(\tau) = \frac{3\tau}{2}\,\big[ 1 + (1-\tau)\,f(\tau) \big] \,,
\end{equation}
with $A_{1/2}(\infty)=1$. For the hypothetical cross section in the denominator of (\ref{eq:sigmaratio}) we use 0.74\,pb, which is the product of the cross section $\sigma=0.157$\,pb at $\sqrt{s}=8$\,TeV quoted in \cite{Heinemeyer:2013tqa} (for $m_h=750$\,GeV) with the boost factor 4.7 accounting for the raise of the gluon luminosity from 8 to 13\,TeV \cite{Franceschini:2015kwy}. We then obtain the values shown in Figure~\ref{fig:production}. With realistic leptoquark masses in the range allowed by collider bounds and capable of explaining the flavor anomalies and reasonable values of the portal coupling we find cross sections large enough to explain the diphoton excess, provided that the branching fraction for the decay $S\to\gamma\gamma$ is of ${\cal O}(1)$.

\section{Diphoton decay}

In the minimal model presented so far, the resonance $S$ can decay either into two gluons or into two photons through leptoquark loops. We find
\begin{equation}
   \frac{\Gamma(S\to\gamma\gamma)}{\Gamma(S\to gg)} 
   = \frac{9 Q_\phi^4}{2 K_{gg}}\,\frac{\alpha^2}{\alpha_s^2} \,,
\end{equation}
with $K_{gg}=1+\frac{\alpha_s}{\pi}\,(\frac{45}{4}+\frac{7}{2} \ln\frac{\mu^2}{M_S^2})\approx 1.33$ for $\mu^2=M_S^2$ \cite{Schroder:2005hy,Chetyrkin:2005ia}. This ratio is roughly of order $3\cdot 10^{-4}$ ($4\cdot 10^{-3}$) for a leptoquark of charge 1/3 (2/3) and thus much too small to explain the signal, in agreement with \cite{Knapen:2015dap}. The situation improves substantially if one introduces new vector-like, color-neutral and electrically charged fermions $\chi$. The existing collider bounds for such particles (assuming not too large electric charges of these states) allow for masses as light as $M_S/2$, and in addition fermion loops are enhanced relative to scalar loops by roughly an order of magnitude. The relevant interaction is
\begin{equation}\label{eq:DMcoup}
   {\cal L} = - \sum_i\,g_{\chi_i S}\,S\,\bar\chi_i\chi_i \,,
\end{equation}
where the sum extends over one or more fermion multiplets $\chi_i$. The neutral partner $\chi_0$ of the lightest lightest multiplet can be stable. The empirical evidence for dark matter provides a strong motivation for introducing such a new fermion. The scalar resonance $S$ would then play the role of a messenger between the visible and dark sectors. Mixing with SM leptons can be avoided either by means of a discrete symmetry or by assigning a charge under a dark gauge group $SU(N_\chi)$ to the new fermions. For general $SU(2)_L$ representations, the couplings in \eqref{eq:DMcoup} induce decays of $S$ into massive electroweak gauge bosons, which are constrained by Run-I data. In Table~\ref{tab:bounds} we compile the relevant bounds and corresponding analyses.

\begin{table}
\centering
\begin{tabular}{ccccc}
\hline
$jj$ & $WW$ & $ZZ$ & $Z\gamma$ \\
\hline
$<2.5$\,pb \cite{CMS:2015neg} & $<40$\,fb \cite{Aad:2015agg} & $<12$\,fb \cite{Aad:2015kna}
 & $<4$\,fb \cite{Aad:2014fha} \\
\hline
\end{tabular}
\caption{\label{tab:bounds}
Bounds (at 95\% CL) on the $pp\to S\to X$ production cross sections obtained in dijet and diboson resonance searches performed in Run-I of the LHC ($\sqrt{s}=8$~TeV).}
\end{table}

We introduce the effective Lagrangian in the unbroken $SU(2)_L\times U(1)_Y$ phase,
\begin{equation}\label{Leff}
\begin{aligned}
   {\cal L}_{\rm eff} 
   &= c_{WW}\,\frac{\alpha}{4\pi s_w^2}\,S\,W_{\mu\nu}^a W^{\mu\nu,a}
    + c_{BB}\,\frac{\alpha}{4\pi c_w^2}\,S\,B_{\mu\nu}B^{\mu\nu} \,,
\end{aligned}
\end{equation}
such that the partial decay width into the various gauge-boson final states can be written as
\begin{equation}\label{eq:gammas}
\begin{aligned}
   \Gamma(S\to\gamma\gamma) 
   &= \frac{\alpha^2 m_S^3}{64\pi^3}\,\left( c_{WW} + c_{BB} \right)^2 , \\
   \Gamma(S\to WW)
   &= \frac{\alpha^2 m_S^3}{32\pi^3}\,\frac{c_{WW}^2}{s_w^4} \\
   &\quad\times \left( 1 - 4x_W + 6x_W^2 \right) \sqrt{1-4x_W} \,, \\
   \Gamma(S\to ZZ)
   &= \frac{\alpha^2 m_S^3}{64\pi^3}
    \left( \frac{c_w^2}{s_w^2}\,c_{WW} + \frac{s_w^2}{c_w^2}\,c_{BB} \right)^2 \\
   &\quad\times \left( 1 - 4x_Z + 6x_Z^2 \right) \sqrt{1-4x_Z} \,, \\
   \Gamma(S\to Z\gamma)
   &= \frac{\alpha^2 m_S^3}{32\pi^3}\!
    \left( \frac{c_w}{s_w}\,c_{WW} - \frac{s_w}{c_w}\,c_{BB} \right)^2\!\!\left( 1 - x_Z \right)^3\!,
\end{aligned}
\end{equation}
where $x_{W,Z}=m_{W,Z}^2/M_S^2$, $s_w=\sin\theta_w$ and $c_w=\cos\theta_w$. If not forbidden by a symmetry, the new fermions can also decay into SM leptons through mixing induced by Yukawa couplings with the Higgs or higher dimensional operators. Such new leptons can be constrained by collider searches, and the authors of \cite{Kumar:2015tna} find a lower bound of $M_\chi>275$\,GeV for the charged component of an electroweak doublet from LHC Run-I data. With 100\,fb$^{-1}$ integrated luminosity at 13\,TeV the bound can be improved to 440\,GeV. For a singlet no bounds are derived, and even for 1000\,fb$^{-1}$ at 13\,TeV an exclusion $M_\chi>200$\,GeV seems optimistic. Higher charges lead to more severe bounds of $360\!-\!460$\,GeV from 8\,TeV data, depending on the coupling structure \cite{Altmannshofer:2013zba}. In the context of the diphoton excess, extra vector-like leptons have also been considered in \cite{Angelescu:2015uiz}. In the following, we consider two scenarios motivated by a connection between the $S\gamma\gamma$ vertex and a dark sector.

\subsection{Triplet Model}

The most minimal implementation of dark matter directly connected to the $S\to\gamma\gamma$ signal is to introduce a single electroweak triplet of Weyl fermions $\chi=\left(\chi_{-1},\chi_0,\chi_1\right)$ with hypercharge~0. We find for the Wilson coefficients in \eqref{Leff}
\begin{equation}
   c_{WW} = \frac23\,\frac{g_{\chi S}}{M_\chi}\,A_{1/2}(\tau_\chi) \,, \qquad
   c_{BB} = 0 \,.
\end{equation}
In this case, the ratios $R_X=\Gamma(S\to X)/\Gamma(S\to\gamma\gamma)$ of the different partial decay widths in \eqref{eq:gammas} are found to be
\begin{equation}\label{eq:Rs}
   R_{WW}\approx 37 \,, \qquad
   R_{ZZ}\approx 11 \,, \qquad
   R_{Z\gamma}\approx 7 \,.
\end{equation}
Fitting the signal of $\sigma(pp\to S\to \gamma\gamma)=4.4$\,fb, while taking into account the boost factor of $4.7$ for gluon-fusion induced production, leads to a tension in all three of these channels, in particular in $S\to Z\gamma$.  From \eqref{eq:Rs} it also follows that the partial diphoton branching ratio is at most
\begin{equation}
   \text{Br}(S\to \gamma\gamma)\lesssim \Big(\sum_{X}R_X\Big)^{-1}\approx 2\% \,.
\end{equation}
Given the production cross section shown in Figure~\ref{fig:production}, this rules out the simplest dark matter embedding in terms of a triplet with hypercharge~0. Note, that the above conclusions are independent of the representation of $\chi$ under $SU(2)_L$, as long as the multiplet carries no hypercharge.

\subsection{Doublet plus Singlet Model}

\begin{figure}
\centering
\includegraphics[width=0.4\textwidth]{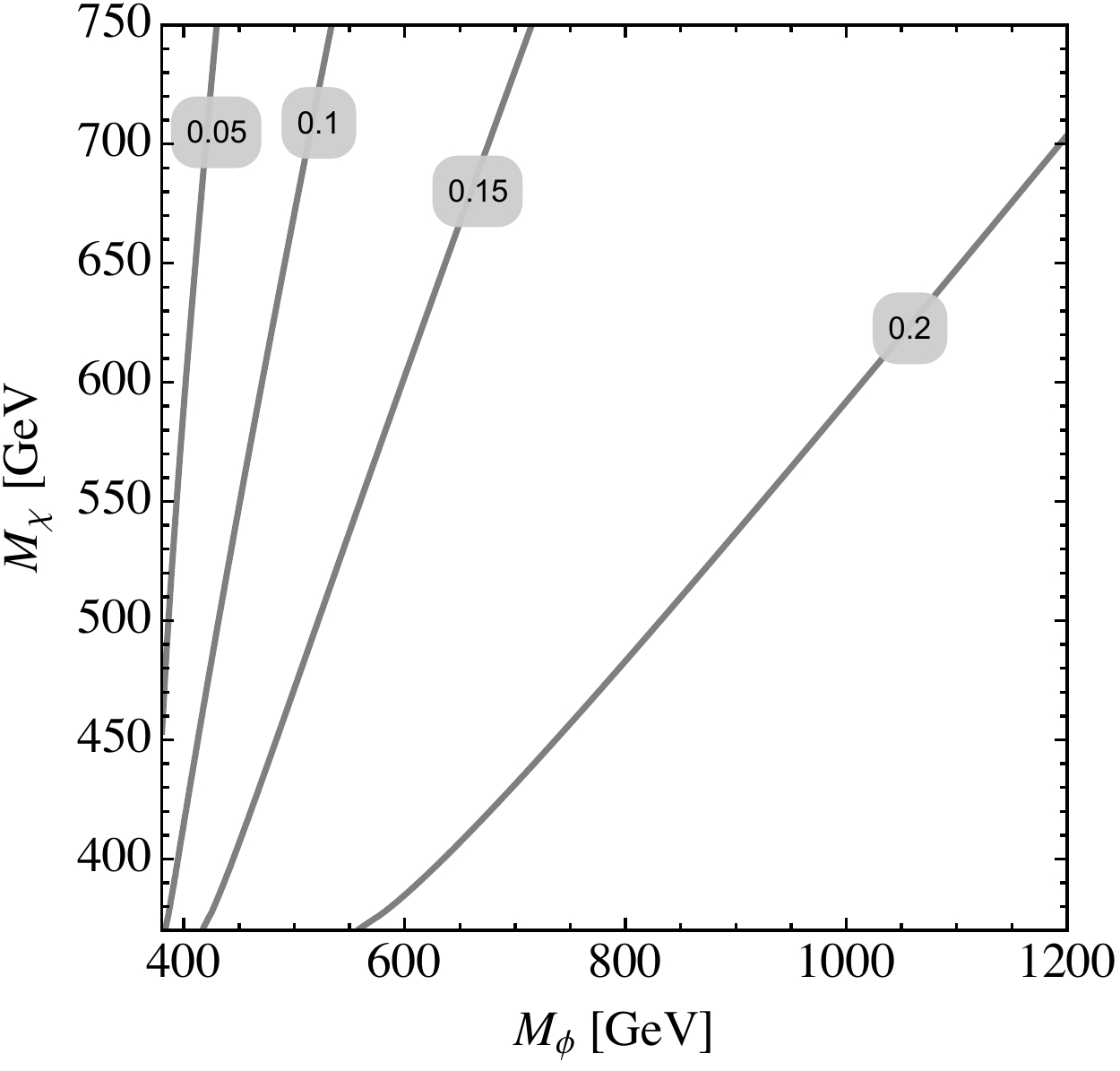}
\vspace{-3mm}
\caption{\label{fig:Brgg} 
Contours of constant branching ratio $\text{Br}(S\to\gamma\gamma)$ for $g_{\phi S}=g_{\chi S}$.}
\end{figure}

\begin{figure}
\centering
\includegraphics[width=0.4\textwidth]{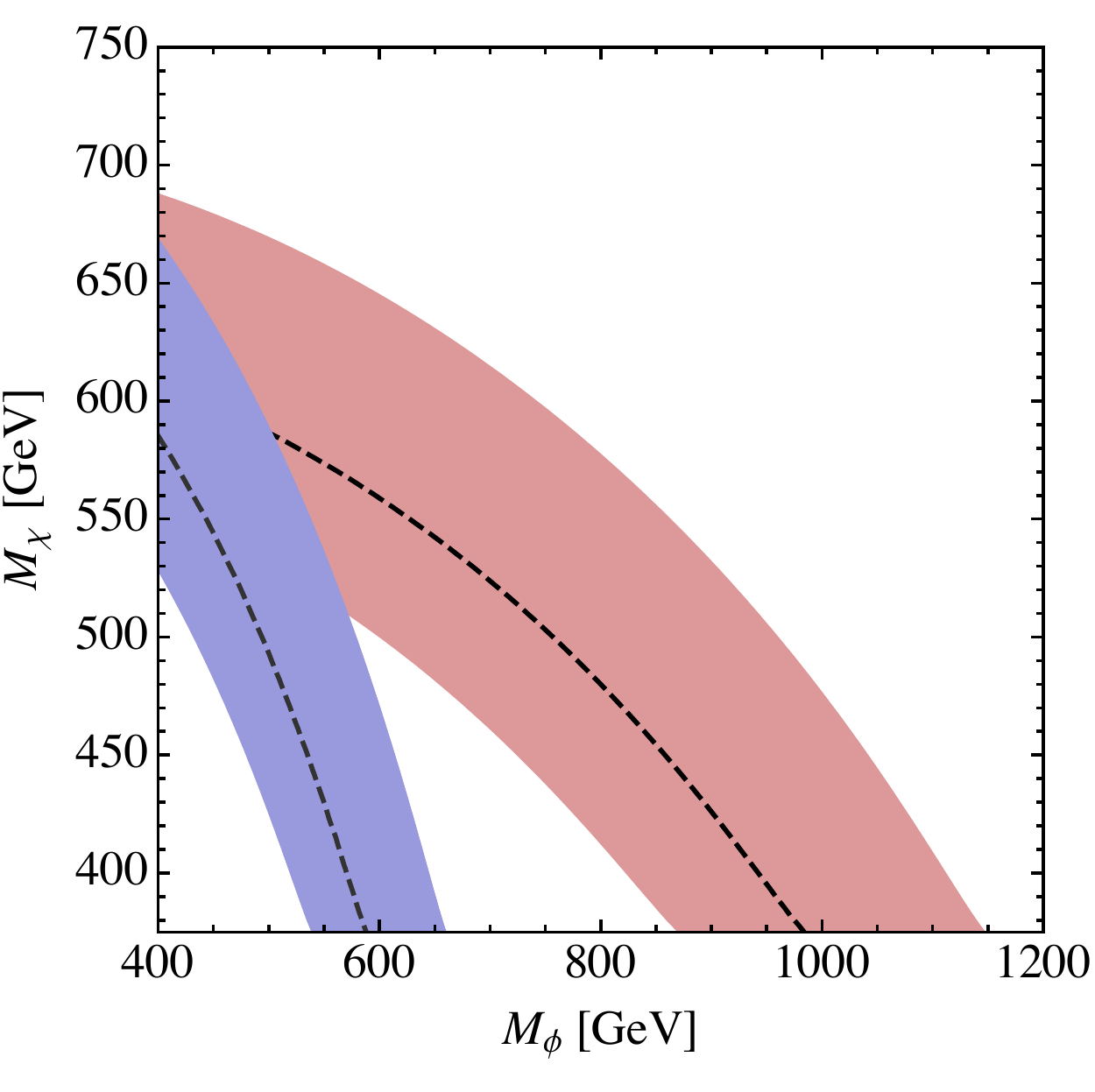}
\vspace{-3mm}
\caption{\label{fig:final1} 
Regions in parameter space reproducing $\sigma(pp\to S)\times\mathrm{Br}(S\to\gamma\gamma)=(4.4\pm 1.1)$\,fb at $1\sigma$ in the two benchmark models with $M_S=750$\,GeV and $g_{\chi S}=2, g_{\phi S}=3$ (blue) and $g_{\chi S}=2, g_{\phi S}=6$ (red). The dashed lines correspond to the central value of the diphoton signal.}
\end{figure}

A doublet with hypercharge~1/2 has a neutral and a charged component and can in principle provide a dark matter candidate as well as an explanation of the diphoton excess. Such a dark matter candidate is, however, excluded by direct detection bounds from $Z$ exchange. We therefore consider an extension with a vector-like $SU(2)_L$ doublet $\psi$ with hypercharge~1/2, a vector-like $SU(2)_L$ singlet $\xi$ with hypercharge~$-1$, and a vector-like $SU(2)_L$ singlet $\eta$ with hypercharge~0. In this model we find \begin{equation}
\begin{aligned}
   c_{WW} &= \frac{1}{3}\,\frac{g_{\psi S}}{M_\psi}\,A_{1/2}(\tau_\psi) \,,\\
   c_{BB} &=\frac{1}{3}\,\frac{g_{\psi S}}{M_\psi}\,A_{1/2}(\tau_\psi)
    + \frac{2}{3}\,\frac{g_{\xi S}}{M_\xi}\,A_{1/2}(\tau_\xi) \,.
\end{aligned}
\end{equation}
For simplicity we will assume that $g_{\chi S}\equiv g_{\psi S}=g_{\xi S}$ and $M_\chi\equiv M_\psi=M_\xi$ in the following, along with $M_\eta < M_\chi$ and $g_{\eta S}$ chosen such that the relic abundance of dark matter is reproduced. In this limit $c_{BB}=3\,c_{WW}$, and one finds
\begin{equation}
   R_{WW}\approx 2.3 \,, \qquad R_{ZZ}\approx 1.1 \,, \qquad R_{Z\gamma}\approx 0.004\,,
\end{equation}
and therefore the diphoton signal can be reproduced without any tension with Run-I measurements. In Figure~\ref{fig:Brgg} we show contours of the branching ratio Br$(S\to \gamma\gamma)$ in the $M_\phi\!-\!M_\chi$ plane for $g_{\phi S}=g_{\chi S}$. A sizable branching ratio can be achieved for a leptoquark mass of $M_\phi>400$\,GeV. Finally, in Figure~\ref{fig:final1} we show a fit to the central value (dashed black line) and the $1\sigma$ uncertainty band of the diphoton excess for $g_{\chi S}=2$ and $g_{\phi S}=3$ (blue) and for $g_{\chi S}=2$ and $g_{\phi S}=6$ (red). For a sizable parameter space the excess can be explained, but values of $M_\phi>1$\,TeV are strongly disfavored. Note, that the benchmarks are chosen to illustrate the preferred parameter space and potential low scale Landau poles can be avoided if multiple leptoquarks are assumed. Further, for $g_{\xi S}>g_{\psi S}$ the branching ratios change in favor of a larger $\text{Br}(S\to\gamma\gamma)$. 

\section{A Dark Matter Sector}

A particularly intriguing scenario emerges if the singlet $S$ not only provides a portal to the dark sector, but also gives mass to these dark fermions through a vacuum expectation value $\langle S\rangle \equiv w/\sqrt{2} $. The couplings \eqref{eq:ourop} and \eqref{eq:DMcoup} are then the consequence of the more fundamental Lagrangian
\begin{equation}\label{eq:DMlag}
   \mathcal{L} = \lambda_{\phi S} S^\dagger S\phi^\dagger\phi
    + g_{\chi S} S\bar\chi^c\epsilon\chi + \mu_S S^\dagger S + \lambda_S (S^\dagger S)^2 \,, ~
\end{equation}
in which $\chi$ denotes an $SU(2)_L\times SU(N_\chi)$ multiplet containing a neutral dark matter candidate $\chi_0$, and $\epsilon$ exchanges members of the multiplet of opposite charge. If the dark matter is part of a vector-like multiplet, the Majorana mass term in \eqref{eq:DMlag} has to be replaced by an appropriate mass term. It follows that $\kappa_{\phi S}=\lambda_{\phi S}\,w$ and $M_\chi= g_{\chi S}\,w/\sqrt{2}$. A discrete remnant of the dark symmetry breaking can forbid an explicit mass term. 

Such a dark matter candidate from a pure electroweak multiplet can be searched for by means of disappearing tracks and monojet searches, for which exclusion limits strongly depend on the $SU(2)_L$ representation. The 14\,TeV LHC is expected to become sensitive to our benchmark model at 3000\,fb$^{-1}$, while an electroweak doublet for example would escape searches for masses $M_\chi>M_S/2$. A future 100\,TeV hadron collider, however, would allow one to discover the dark matter candidate in both cases \cite{Low:2014cba}. A comprehensive analysis of the dark matter phenomenology is left for future work.

\section{\boldmath $S\to\tau^+\tau^-\!$ and $S\to t \bar t$ decay modes}

\begin{figure}[t]
\includegraphics[width=0.45\textwidth]{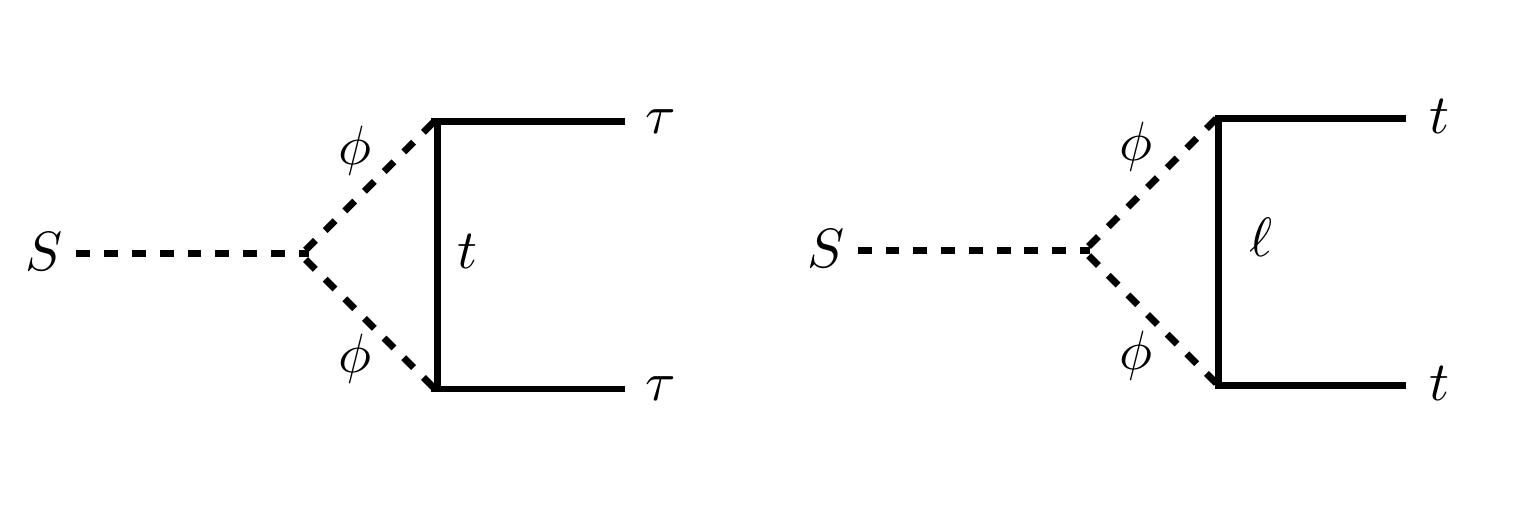}
\caption{Diagrams contributing to the exotic decay channel $S\to\tau^+\tau^-$ (left) and $S\to t\bar t$ (right). The first mode is characteristic for our model.\label{fig:LeptDecay}}
\end{figure}

A key signature of models in which the new resonance $S$ is produced in gluon fusion via leptoquark loops may consist in exotic decays of $S$ into lepton pairs via the diagram shown in Figure~\ref{fig:LeptDecay}. This process is chirally suppressed by the mass of the SM fermion in the loop, and hence significant rates can only arise if this fermion is a top quark. For the leptoquarks studied in \cite{Hiller:2014yaa} the top-quark only couples to neutrinos, yielding invisible decays. For the leptoquark studied in \cite{Bauer:2015knc}, on the other hand, and the coupling structure preferred by the explanation of flavor anomalies proposed in this paper, the decays into tau pairs are preferred. Neglecting the tau lepton mass and working at leading order in the ratio $m_t^2/M_\phi^2$, we find
\begin{equation}
   \frac{\Gamma(S\to\tau^+\tau^-)}{\Gamma(S\to gg)} 
   \approx \frac{81}{4 K_{gg}}\,\frac{m_t^2}{M_S^2}\,
    \frac{\big|\lambda_{t\tau}^L\,\lambda_{t\tau}^{R\ast}\big|^2}{\alpha_s^2\,\pi^2}\,
    \bigg| \frac{\tau_\phi\,f(\tau_\phi)}{A_0(\tau_\phi)} \bigg|^2 \,,
\end{equation}
where $\lambda_{t\tau}^{L,R}$ denote the relevant couplings of the leptoquark. Numerically, we find that this ratio varies between 5.8 and 6.8 times $|\lambda_{t\tau}^L\,\lambda_{t\tau}^{R\ast}|^2$ for $M_\phi$ between 0.6 and 1.2\,TeV, where we have used $m_t\equiv m_t(M_S)\approx 147$\,GeV. The phenomenological analysis in \cite{Bauer:2015knc} shows that the generation-diagonal left-handed leptoquark couplings are preferentially of ${\cal O}(1)$, while the right-handed couplings need to be suppressed. For the second-generation couplings one finds that $|\lambda_{c\mu}^L\,\lambda_{c\mu}^{R\ast}|\approx 0.1$ provides a good fit to the data. If a similar value is assumed for the product of the third-generation couplings, one would expect that the $S\to\tau^+\tau^-$ branching ratio can be close to 10\% of the $S\to gg$ branching ratio. Observing these exotic decays would provide a smoking-gun signature of our model and a dedicated search provides a worthwhile target for a future high-energy hadron collider. 

The decay of $S$ into top-quark pairs is induced by the second diagram in Figure~\ref{fig:LeptDecay}, where a sum over lepton flavors $\ell=e,\mu,\tau$ is implied. In this case, we find 
\begin{equation}\label{eq:ttrat}
   \frac{\Gamma(S\to t \bar t)}{\Gamma(S\to gg)} 
   \approx \frac{27}{64 K_{gg}} \frac{m_t^2}{M_S^2} 
    \frac{\big|\lambda_{t\ell}^L\,\lambda_{t\ell}^{L\ast}\big|^2}{\alpha_s^2\,\pi^2} 
    \bigg| \frac{\tau_\phi\,\tilde f(\tau_\phi)}{A_0(\tau_\phi)} \bigg|^2
    \!\left(1-\tau_t\right)^\frac{3}{2} \!,
\end{equation}
in which $\tilde f (\tau)= 1-2\sqrt{\tau-1}\arcsin (\frac{1}{\sqrt{\tau}})+\tau\arcsin^2(\frac{1}{\sqrt{\tau}})$, and we use the pole mass $m_t\approx 173$\,GeV in the phase-space factor involving $\tau_t=4m_t^2/M_S^2$. If right-handed leptoquark couplings are also included, one must replace $\lambda_{t\ell}^L\,\lambda_{t\ell}^{L\ast}\to\lambda_{t\ell}^L\,\lambda_{t\ell}^{L\ast}+\lambda_{t\ell}^R\,\lambda_{t\ell}^{R\ast}$; however, the additional contribution is negligible due to a suppression by a factor $\sim 10^{-2}$ with respect to the leading term, which is a generic feature of the model proposed in \cite{Bauer:2015knc}. Numerically, the ratio \eqref{eq:ttrat} varies between 0.09 and 0.10 times $|\lambda_{t\ell}^L\,\lambda_{t\ell}^{L\ast}|^2$ for $M_\phi$ between 0.6 and 1.2\,TeV. The generation-diagonal left-handed leptoquark couplings are preferentially of ${\cal O}(1)$ \cite{Bauer:2015knc}. We find that the branching ratio $S\to t\bar t$ varies between $2\%-\!10\%$ of the $S\to gg$ branching ratio for $|\lambda_{t\ell}^L\,\lambda_{t\ell}^{L\ast}|=0.5-1$. For the benchmarks shown in Figure~\ref{fig:final1} this changes the fit for $M_\phi\lesssim M_\chi$ and only by a small amount. Finally, we note that for a leptoquark mass of $M_S/2<M_\phi<M_S$, tree-level decays such as $S\to\phi\phi^*\to t\bar t\,\tau^+\tau^-$ can contribute with branching ratios comparable to $S\to\tau^+\tau^-$ and would provide an additional non-trivial test of our model.

\section{Conclusions}

We have shown that color-triplet scalar leptoquarks with mass in the TeV range, which have been invoked in the past to explain some of the persisting anomalies in the flavor sector seen at the LHC and the $B$ factories, can account for the recently observed excess in the diphoton production rate seen in the first $\sqrt{s}=13$\,TeV run at ATLAS and CMS. With a leptoquark mass in the range near or below 1\,TeV and a natural value of the portal coupling to a new scalar resonance $S$, the production cross section $\sigma(pp\to S)$ via leptoquark-induced gluon fusion is predicted to lie between a few and a few tens of femtobarn. A large branching ratio for the diphoton decay mode of the resonance $S$ can be obtained if this resonance is coupled to a multiplet of new color-neutral charged fermions $\chi$ with weak-scale masses. While the electrically charged members of this multiplet mediate the decay $S\to\gamma\gamma$, its neutral component $\chi_0$ can be stable and provides a good candidate for WIMP dark matter. Our model predicts that the resonance should have an exotic decay mode into pairs of tau leptons, with a branching ratio that could reach 10\% under reasonable assumptions.

The model described here is a prototype of a more general scenario, in which the scalar resonance $S$ acts as a mediator between the SM and a new sector containing a dark-matter state $\chi_0$ as a member of a color-neutral vector-like fermion multiplet. While $S$ is produced in gluon fusion via loops containing very heavy new colored particles -- the scalar leptoquark $\phi$ in our model -- its diphoton decay is predominantly mediated by loops containing lighter, electrically charged fermion states of the multiplet $\chi$. The production via a heavy scalar implies that the dijet decay model $S\to gg$ has a suppressed width, and hence existing dijet bounds can readily be avoided. As a result, in our class of models the total width of the new resonance $S$ is very small, typically $\Gamma_S\sim 1$\,MeV.

\vspace{2mm}\noindent
M.B.\ acknowledges the support of the Alexander von Humboldt Foundation. M.N.\ is supported by the ERC Advanced Grant EFT4LHC, the PRISMA Cluster of Excellence EXC 1098 and grant 05H12UME of the German Federal Ministry for Education and Research.

\end{document}